\documentclass[useAMS,usenatbib,usegraphicx]{mn2e}
\usepackage{amsmath}
\usepackage{amssymb}
\usepackage{graphicx}
\usepackage{latexsym}
\usepackage{aas_macros}
\usepackage{natbib}
\usepackage[squaren]{SIunits}
\usepackage{array}
\usepackage{tabularx}

\title[Bondi--Hoyle Flow with Radiative Feedback]{The Effect of Radiative Feedback on Bondi--Hoyle Flow around a Massive Star}
\author[R.G.~Edgar \& C.J.~Clarke]{Richard~Edgar$^1$\thanks{Email: rge21@ast.cam.ac.uk}
 and Cathie~Clarke$^1$ \\
$^1$Institute of Astronomy, Madingley Road, Cambridge CB3 0HA, Great Britain} 
\date{\today}

\pagerange{\pageref{firstpage}--\pageref{lastpage}} \pubyear{2003}

\newcommand{\scinot}[2]{\ensuremath{#1 \times 10^{#2}}}

\newcommand{\rdust}[1]{\ensuremath{r_{\textrm{d}}^{#1}}}
\newcommand{\tauR}{\ensuremath{\tau_{\textrm{R}}}}

\newcommand{\vinf}[1]{v_{\infty}^{#1}}
\newcommand{\rhoinf}[1]{\rho_{\infty}^{#1}}
\newcommand{\cinf}[1]{c_{\infty}^{#1}}

\newcommand{\zetaHL}[1]{\zeta_{\text{HL}}^{#1}}
\newcommand{\mdotHL}[1]{\dot{M}_{\text{HL}}^{#1}}

\newcommand{\zetaBH}[1]{\zeta_{\text{BH}}^{#1}}
\newcommand{\mdotBH}[1]{\dot{M}_{\text{BH}}^{#1}}

\newcommand{\rhodust}[1]{\ensuremath{\rho_{\textrm{d}}^{#1}}}
\newcommand{\vdust}[1]{\ensuremath{v_{\textrm{d}}^{#1}}}
\newcommand{\rhoinside}[1]{\ensuremath{\rho_{\textrm{i}}^{#1}}}
\newcommand{\vinside}[1]{\ensuremath{v_{\textrm{i}}^{#1}}}

\newcommand{\thetac}[1]{\ensuremath{\theta_{\textrm{c}}^{#1}}}

\newcommand{\HII}{H{\sc ii}}   

\newcommand{\Zeus}{{\sc Zeus2d}}

\newcommand{\differd}[1]{\textrm{d}^{#1}}
\newcommand{\differ}[1]{\differd{}#1}
\newcommand{\sdiffer}[1]{\, \differ{#1}}

\addunit{\cm}{\centi\metre}

\addunit{\percubiccmnp}{\cm \rpcubed}
\addunit{\grampercubiccmnp}{\gram \usk \percubiccmnp}

\addunit{\cmpersecnp}{\cm \usk \reciprocal \second}

\addunit{\yyear}{yr} 

\newcommand{\mySun}{\odot}

\addunit{\Msol}{\ensuremath{\mathrm{M}_{\mySun}}}
\addunit{\Lsol}{\ensuremath{\mathrm{L}_{\mySun}}}

\begin{document}

\label{firstpage}

\maketitle

\begin{abstract}
We apply an algorithm for radiative feedback on a dusty flow
(detailed in \citet{2003MNRAS.338..962E}) to the problem of
Bondi--Hoyle accretion.
This calculation is potentially relevant to the
formation of massive stars  in ultradense cores of stellar clusters.
We find that radiative feedback is \emph{more effective} than in
the case of previous calculations in spherical symmetry.
The Bondi-Hoyle geometry implies that 
material is flowing nearly tangentially when it experiences the sharp
radiative impulse at the dust destruction radius, and consequently
it is readily perturbed into outflowing orbits.
We find that it is difficult for stellar masses to grow beyond around
\unit{10}{\Msol} (for standard interstellar dust abundances).
We discuss the possible implications of this result for the formation
mechanism of OB stars in cluster cores.
We end by proposing a series of conditions which must be fulfilled if
Bondi--Hoyle accretion is to continue.
\end{abstract}



\section{Introduction}

Massive stars are immensely important astrophysical objects,
far more so than their rarity would imply.
They are incredibly bright, being visible at cosmological
distances.
They manufacture most of the metals in the Universe, both
during their brief lives, and in the supernov\ae{} which
mark their deaths.
Massive stars also exert large forces on their surroundings,
through winds, ionisation and radiation pressure.

Despite this importance, massive star formation is a poorly
understood process.
Observational studies are hampered by the distance to massive
star forming regions, and the high degree of obscuration in
such regions.
From a theoretical point of view, the very existence of
massive stars presents a challenge.
Early studies (in spherical symmetry) suggested that
radiation pressure on dust should prevent the assembly of such
stars unless the opacity of the dust is arbitrarily reduced
by an order of magnitude compared with interstellar values
\citep{1986ApJ...310..207W,1987ApJ...319..850W}
More recently, simulations under conditions of axisymmetry
have shown that accretion via a disc can ease the problem
somewhat \citep{2002ApJ...569..846Y}.
However, other authors have continued to seek other possible
formation mechanisms for massive stars.
One suggestion (postulated by \citet{1998MNRAS.298...93B} and
developed as \citet{2002MNRAS.336..659B}) is that massive
stars are formed in the ultra-dense cores of young stellar
clusters.
Such cores are associated with OB star formation
\citep{2002hsw..work..283B}, although observed cores do not
reach the densities required by \citeauthor{1998MNRAS.298...93B}.
In the merger models of \citeauthor{1998MNRAS.298...93B},
continued accretion onto the stars drives an adiabatic
contraction of the cluster core to the high densities
required.
\citet{2001MNRAS.323..785B} used hydrodynamical simulations
of cluster formation to show that accretion onto the stars
in the core is reasonably well described by Bondi--Hoyle
accretion (i.e. the accretion cross section is appropriate
to a compact object moving supersonically through a
gaseous medium).
However, such simulations include no feedback, and
thus cannot address the question of whether massive stars
can continue to accrete by this mechanism in an
ultra-dense cluster core.

The interaction of outflowing radiation with a dusty accretion flow
may be envisaged in the following terms:
Dust sublimes at a radius in the flow, $\rdust{}$, where its temperature
attains a value of \unit{1700}{\kelvin}.
Inward of this radius, there is no significant momentum exchange between the
radiation field and the gas, until very small radii, where the gas become
ionised.
Radiation that impacts the dusty flow at $\rdust{}$ is absorbed within a
very narrow region, since the Rosseland mean opacity of the dust at the
stellar radiation temperature  ($\sim \unit{30000}{\kelvin}$) is very
high.
The absorption of stellar luminosity $L$ in this region is associated with
momentum transfer to the gas at a rate $L/c$ and the inflow suffers a
sharp deceleration at this point. 
This absorbed radiation is subsequently thermalised, i.e.
re-emitted at a temperature equal to the local dust temperature
($\sim \unit{1700}{\kelvin}$).
The  rate of momentum transfer to the flow that is associated with this
thermalised field is $\sim \tauR{} L/c$, where
$\tauR{}$ is the optical depth of the flow to the thermalised
radiation.

In this paper, we explore the issue of radiative feedback
by radiation pressure on dust in the Bondi--Hoyle geometry.
In Bondi--Hoyle accretion of isothermal gas,
the flow is gravitationally focused behind the massive object (star) 
and forms a shock in the downstream direction.
For gas entering with an impact parameter less than a critical value
(the Bondi--Hoyle radius), the loss of tangential momentum renders
the shocked gas bound to the star and enables it to be accreted by
way of radial infall along the shock.
Such accretion column geometry evidently has different
implications for the interaction between the mass flow and radiation
field than either of the geometries in which radiative feedback
has been simulated to date (i.e. spherical or disc accretion).
Our study will be a first look at how radiative feedback on dust
operates in Bondi--Hoyle geometry
We will also be mindful that in the environment in which such
accretion may be important, it is not only the geometry that is
different from conventional studies but also the physical
parameters.
In particular, the hydrodynamical simulations suggest
that the mean gas densities in the hypothesised ultra-dense core
phase is extremely high (\unit{10^6-10^9}{\percubiccmnp}) and we
will find that this is an important factor in determining the
efficacy of radiative feedback.

In the simulations described below, we employ an algorithm for
simulating the effect of radiative feedback on a dusty flow that
we developed in a previous paper \citep{2003MNRAS.338..962E}.
The aim of this algorithm is to preserve the sharp deceleration
at the point where the dust sublimes, while avoiding the computational
cost of a full frequency dependent calculation.
\citet{1987ApJ...319..850W} showed that a frequency averaged (grey)
approach missed this sharp impulse (which corresponds to $L/c$ worth
of momentum).
The reader should refer to \citet{2003MNRAS.338..962E} for a full
description (and test) of our method, but the general algorithm is
as follows:
\begin{itemize}
\item The radiation field is split into direct and thermalised
      components
\item The direct field is attenuated using wavelength dependent
      opacities
\item The balance of the energy is placed into the thermalised
      field, which is then solved using the diffusion approximation
\end{itemize}
The initial location of the dust destruction front is estimated
based on the temperature of a bare grain in the stellar radiation
field, but the algorithm iterates if the thermalised field implies
the dust melts further from the star.
The mechanical effect of anisotropic scattering is included in
the radiation pressure opacity, but the scattered radiation is
not followed further.
In this paper, we solve the diffusion approximation along each
radius under the assumption of spherical symmetry.
This was done for reasons of simplicity, although it should be possible
to relax this approximation in future work.
Spherical symmetry is obviously not a good approximation for
the accretion column, but we shall show that the flow is disrupted
in a region where the deviations from spherical symmetry are not
large.
This gives us some confidence that our conclusions are not merely
an artifact of our algorithm.

The structure of this paper is as follows:
In section~\ref{sec:basics} we present a brief mathematical analysis
of Bondi--Hoyle flow.
Section~\ref{sec:Params} details our search for suitable simulation
parameters.
The results from our hydrodynamic simulations follow in
section~\ref{sec:dynamics}
We discuss our results and present out conclusions in
sections~\ref{sec:discussion} and~\ref{sec:conclude} respectively.


\section{Bondi--Hoyle Accretion: Analytic Approach}
\label{sec:basics}

\begin{figure}
\centering
\includegraphics[scale=0.7]{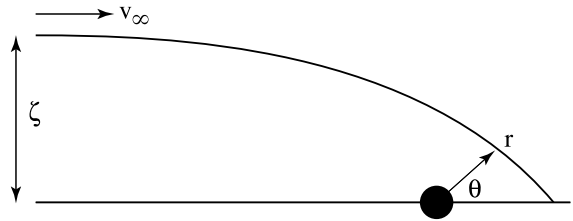}
\caption{Sketch of the Bondi--Hoyle accretion geometry}
\label{fig:BHgeometry}
\end{figure}

\citet{1939PCPS.34..405} considered accretion by a star (mass $M$) moving at
a steady speed ($\vinf{}$) through an infinite gas cloud.
The gravity of the star focuses the flow into a wake which it then
accretes.
The geometry is sketched in figure~\ref{fig:BHgeometry}.

By considering ballistic orbits, \citeauthor{1939PCPS.34..405}
concluded that all material with an impact parameter satisfying
\begin{equation}
\zeta < \zetaHL{} = \frac{2 G M}{\vinf{2}}
\label{eq:HoyleLyttletonRadiusDefine}
\end{equation}
would be accreted.
If the cloud has density $\rhoinf{}$, the mass flux is
\begin{equation}
\mdotHL{} = \pi \zetaHL{2} \vinf{} \rhoinf{} = \frac{4 \pi G^2 M^2 \rhoinf{}}{\vinf{3}}
\label{eq:HoyleLyttletonAccRateDefine}
\end{equation}
which is known as the Hoyle--Lyttleton accretion rate.
By considering the stability of the accretion column
(the wake following the point mass on the $\theta = 0$ axis),
\citet{1944MNRAS.104..273B} concluded that the true accretion
rate could be as little as half this value.

\citet{1952MNRAS.112..195B} studied spherically symmetric
accretion onto a point mass.
The analysis shows (see e.g. \citet{2002apa..book.....F})
that a Bondi radius may be defined as
\begin{equation}
r_{\text{B}} = \frac{2GM}{c_{\text{sB}}^{2}}
\label{eq:BondiRadiusDefine}
\end{equation}
where $c_{\text{sB}}$ is the sound speed at $r_{\text{B}}$.
Flow outside this radius is subsonic, and the density
is almost uniform.
Within it, the gas becomes supersonic and moves towards a freefall
solution.
The similarities between equations~\ref{eq:HoyleLyttletonRadiusDefine}
and~\ref{eq:BondiRadiusDefine}
led \citeauthor{1952MNRAS.112..195B} to propose an interpolation
formula:
\begin{equation}
\dot{M} = \frac{2 \pi G^2 M^2 \rhoinf{}}{(\cinf{2} + \vinf{2})^{3/2}}
\label{eq:BondiHoyleAccRateInterpolate}
\end{equation}
This is known as the Bondi--Hoyle accretion rate
The corresponding $\zetaBH{}$ is formed by analogy with
equation~\ref{eq:HoyleLyttletonAccRateDefine}.
On the basis of their numerical calculations, \citet{1985MNRAS.217..367S}
suggest that equation~\ref{eq:BondiHoyleAccRateInterpolate} should
acquire an extra factor of two, to become
\begin{equation}
\mdotBH{} = \frac{4 \pi G^2 M^2 \rhoinf{}}{(\cinf{2} + \vinf{2})^{3/2}}
\label{eq:BondiHoyleAccRateDefine}
\end{equation}
which then matches the original Hoyle--Lyttleton rate as the sound
speed becomes insignificant.

\citet{1979SvA....23..201B} derived the following equations to describe
the flow, based on the ballistic approximation:
\begin{eqnarray}
v_{r} & = & - \sqrt{ \vinf{2} + \frac{2 G M}{r} - \frac{\zeta^2 \vinf{2}}{r^2}}
\label{eq:BHanalyticvr} \\
v_{\theta} & = & \frac{\zeta \vinf{}}{r}
\label{eq:BHanalyticvtheta} \\
r & = & \frac{ \zeta^2 \vinf{2}}{GM(1+\cos \theta) + \zeta \vinf{2} \sin \theta}
\label{eq:BHanalyticr} \\
\rho & = & \frac{\rhoinf{} \zeta^2}{r \sin \theta ( 2 \zeta - r \sin \theta )}
\label{eq:BHanalyticrho}
\end{eqnarray}
The first three equations are obtained from the Newtonian
orbits.
To obtain the final equation, one must solve the gas continuity equation
using the previous three.

\subsection{Simple Feedback}

The simplest way of including radiative feedback is
to place $L/c$ worth of momentum into the flow at the
dust destruction radius, $\rdust{}$.
This misses the effect of the thermalised radiation field,
but is a useful first approximation.

\begin{figure}
\centering
\includegraphics[scale=0.7]{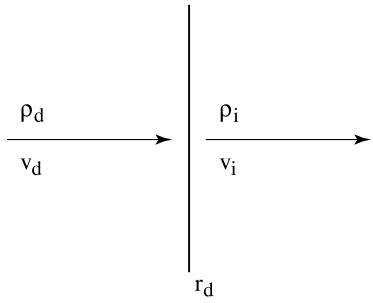}
\caption{Sketch of conditions at the dust destruction radius}
\label{fig:BHshockJumpDiag}
\end{figure}

Consider the flow sketched in figure~\ref{fig:BHshockJumpDiag}.
The `d' subscripts denote quantities immediately outside $\rdust{}$,
`i' subscripts denote values immediately inside $\rdust{}$.
All the velocities are radial, since the radiative impulse is
radial.
The $\theta$ velocities will be unchanged.
The flow is incident from the left, encounters the dust destruction front,
and changes density and velocity.
Assuming that the gas remains isothermal, we can conserve mass and
momentum to find
\begin{eqnarray}
\rhodust{} \vdust{} & = & \rhoinside{} \vinside{} \\
2 \pi \rhodust{} \vdust{2} \rdust{2} \sin \theta \sdiffer{\theta} - \frac{L}{2c} \sin \theta \sdiffer{\theta}
& = &
2 \pi \rhoinside{} \vinside{2} \rdust{2} \sin \theta \sdiffer{\theta}
\end{eqnarray}
where the $\sin \theta \sdiffer{\theta}$ terms are taking care of solid
angles.
This pair of equations bears a suspicious resemblance to the Rankine--Hugoniot
conditions for shocks.
We can eliminate the value of $\rhoinside{}$ between these equations, to give
\begin{equation}
\vinside{} = \vdust{} - \frac{L}{4 \pi \rdust{2} c \rhodust{} \vdust{}}
\label{eq:BHshockVelChange}
\end{equation}
At first sight, this appears to be independent of angle.
However, it is not - $\vdust{}$ and $\rhodust{}$ are somewhat
complex functions of $\theta$.

In simple physical terms, the flow will slow dramatically
in the radial direction as it passes through $\rdust{}$.
This will cause gas streamlines to move onto more circular orbits
(since their angular momentum is unaffected).
The gas streams could then flow around inside $\rdust{}$, and still
end up on the accretion column trailing the star.
Indeed, if $\rdust{}$ were larger than $\zetaBH{}$, then a small
radiative impulse could \emph{increase} the accretion rate.
Material originating from large $\zeta$ would be moved onto orbits
closer to circular (since the \emph{radial} impulse would not affect
angular momentum), and hence be more likely to be bound to the
star after encountering the accretion column.
For a protostar in a protocluster, this would require
$\vinf{} > \unit{36}{\kilo\metre\usk\reciprocal\second}$, which is a
little on the high side.
The accretion rate (proportional to $\vinf{-3}$ - cf
equation~\ref{eq:HoyleLyttletonAccRateDefine}) would also be rather
low, unless the density were increased to compensate.

There will be a gas streamline which has its closest approach to the
star at $\rdust{}$.
Equation~\ref{eq:BHshockVelChange} is obviously not going to apply to
this streamline, since it will have $\vdust{} = 0$.
If the angle, $\thetac{}$, at which this occurs is close enough to
zero, this will not be a problem, since it will be lost in the wake.
However, there will be a range of angles with $\theta < \thetac{}$
for which equation~\ref{eq:BHshockVelChange} will imply a negative
value for $\vinside{}$.
Physically, this corresponds to the flow being turned around at
$\rdust{}$.
While this is fine for particles, it will be disastrous for the
ballistic approximation of the gas.
Streamlines will cross, destroying the assumptions necessary to
treat the gas using simple Newtonian theory.

Using equation~\ref{eq:BHshockVelChange} and the fact that the
velocity in the $\theta$ direction is unchanged, it is fairly
straightforward to compute the trajectory followed by the
gas inside the $\rdust{}$ (on the assumption of ballistic flow).

\begin{figure}
\centering
\includegraphics[scale=0.4]{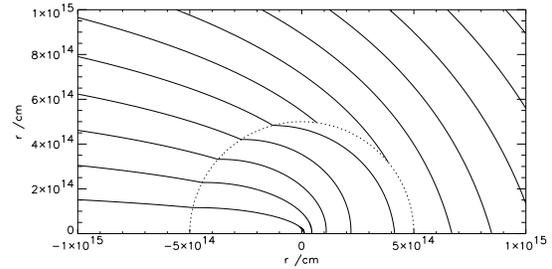}
\caption{Ballistic trajectories in Bondi--Hoyle flow.
The dust destruction front is marked by the dotted line}
\label{fig:BHanalyticTrajectories}
\end{figure}

Figure~\ref{fig:BHanalyticTrajectories} plots some sample trajectories
for flow past a \unit{10}{\Msol} star.
In this plot, $\rhoinf{} = \unit{10^{-16}}{\grampercubiccmnp}$ and
$\vinf{} = \unit{\scinot{5}{5}}{\cmpersecnp}$.
We shall show in the next section that these
parameters are appropriate for the ultra-dense cores postulated by
\citeauthor{2002MNRAS.336..659B}.
To make the perturbations to the flow visible, we used an artificially
large value for the luminosity of \unit{10^4}{\Lsol} (approximately
twice the ZAMS value for a \unit{10}{\Msol} star).
We placed the dust destruction radius at
$\rdust{} = \unit{\scinot{5}{14}}{\centi\metre}$, which is a reasonable
value for these conditions.

Our simplified feedback model implies that radiation pressure on dust
will give rise to orbit crossing (and hence the formation of a shock cone)
in the downstream direction (i.e. for $\theta < \thetac{}$).
For a luminosity appropriate to a \unit{10}{\Msol} ZAMS star,
$\thetac{}  \approx \unit{0.2 \pi}{\radian}$, which far exceeds the width
of the unperturbed Bondi--Hoyle wake.

To proceed further, it is necessary to resort to full hydrodynamic
calculations, both because these can follow the flow in the orbit crossing
regime and because they can include additional feedback from the thermalised
radiation field.
This important effect, omitted in the above analysis, decelerates the
flow before it encounters the dust destruction front.
In addition, we can use hydrodynamical simulations to explore the case
that the equation of state of the gas is not isothermal, since it is
well known (e.g. \citet{1971MNRAS.154..141H,1979MNRAS.188...83H}
and \citet{1994ApJ...427..342R,1994ApJ...427..351R}) that Bondi--Hoyle flow
is not well described by the equations of \citeauthor{1979SvA....23..201B}
in this case.


\section{Parameters}
\label{sec:Params}

Simulating Bondi--Hoyle flow in the context of star formation
is rather difficult, due to the large range in length scales
which must be resolved.
Although we are interested in stars rather more massive than
the compact objects typically studied in Bondi--Hoyle simulations,
the protostellar velocities in a cloud core are generally much
lower as well - typically only a few
$\kilo\metre\usk\reciprocal\second$.
This tends to make $\zetaHL{}$ rather large
(cf equation~\ref{eq:HoyleLyttletonRadiusDefine}),
and many times larger than the protostellar radius.
We can reduce the problem, since we are not interested in following
the flow right down to the protostar, but only to the dust destruction
front, $\rdust{}$.\footnote{$\rdust{}$ is calculated by computing the
sublimation temperature of a dust grain in the stellar radiation field.
See \citet{2003MNRAS.338..962E} for details}
However, the ratio $\zetaHL{}/\rdust{}$ can still be painfully large.
Another motivation for keeping $\zetaHL{}$ small is self-gravity.
If the mass enclosed in the simulation is comparable to that of the
protostar, consistency requires that the flow is self-gravitating - which
would require a great deal of computational effort.
The \HII{} region surrounding the protostar must also be kept away from
the dust.
If ionised gas reaches the dust, then it might be able to destroy the
dust before the grains reach their nominal sublimation temperature.

To find suitable simulation parameters ($M$, $\vinf{}$, $\rhoinf{}$),
we make use of the analytic formul\ae{} for pressure free Bondi--Hoyle
flow derived by \citet{1979SvA....23..201B}.
We estimate the volume of ionised gas by balancing the number of
ionisations and recombinations along each sight line, in a
straightforward generalisation of the Str\"{o}mgren Sphere
\citep{1939ApJ....89..526S}.
To calculate the number of ionising photons, we make use of the
ZAMS formul\ae{} of \citet{1996MNRAS.281..257T} and the
Stefan--Boltzmann Law.
We seek parameters which satisfy the following conditions:
\begin{itemize}
\item The predicted accretion rate will cause a significant
      change in mass on a timescale of a million years or so
\item The volume of ionised gas must lie within $\rdust{}$
\item The ratio $\rdust{}/\zetaBH{}$ must not
      be too small
\end{itemize}
Note that the first two requirements are motivated by physical
considerations.
The third is a computational practicality.
Unfortunately, these requirements tend to conflict with
each other - a faster gas flow will help reduce $\zetaBH{}$,
but it will also reduce the central concentration of the
fluid.
This will let the ionised region grow.

Based on information supplied by Ian Bonnell (published as
\citet{2002MNRAS.336..659B}) the reasonable ranges for the relevant
parameters are given in Table~\ref{tbl:PossParamsBHcluster}.
However, the upper end of each range is beginning to push limits
to the extreme.
In particular, there is little point seeking Bondi--Hoyle accretion
solutions for a \unit{100}{\Msol} star, unless we have already
established that it can grow to this mass.
Having said this, more massive stars make it easier to keep the ionisation
problem to a minimum.
This is because $\rdust{}$ is an `area' effect (cf the Stefan--Boltzmann Law)
whereas ionisation is a `volume' effect (cf the Str\"{o}mgren Sphere).

\begin{table}
\centering
\begin{tabular}{l|r@{$\,<\,$}c@{$\,<\,$}l}
Gas Density  & \unit{10^{-18}}{\grampercubiccmnp}  &  $\rhoinf{}$  &  \unit{10^{-15}}{\grampercubiccmnp} \\
Gas Velocity & \unit{10^{4}}{\cmpersecnp}          &  $\vinf{}$    &  \unit{10^{7}}{\cmpersecnp} \\
Stellar Mass & \unit{10}{\Msol}                    &  $M$          &  \unit{100}{\Msol}
\end{tabular}
\caption{Range of Parameters for an ultra--dense core \citep{2002MNRAS.336..659B}}
\label{tbl:PossParamsBHcluster}
\end{table}

From this, the parameters given in Table~\ref{tbl:ChoseParamsBHsims} squeeze through
the selection criteria.
These give an accretion rate of \unit{\scinot{3}{-4}}{\Msol\usk\reciprocal\yyear}
and $\zetaBH{} = \unit{\scinot{1.1}{16}}{\centi\metre}$.
The dust destruction front due to direct stellar radiation is located at
$\rdust{} = \unit{\scinot{2.1}{14}}{\centi\metre}$.
The ionised gas should always lie within $\sim \unit{\scinot{1.3}{14}}{\centi\metre}$ of the
star,\footnote{Note that
this is still substantially smaller than the Hypercompact \HII{} regions discussed
by \citet{2002hsw..work...81K}} which is sufficiently inside $\rdust{}$ for our purposes.
This is helped by the following considerations:
\begin{itemize}
\item The value of $\rdust{}$ is likely to increase, due to the thermalised
      radiation field
\item The escape velocity at the edge of the \HII{} region is on the order of
      forty kilometres per second, while the sound speed in the ionised gas
      is only a few kilometres per second
\end{itemize}
Therefore, ionisation is unlikely to be a problem at the start of the simulations.
So long as accretion continues, the \HII{} region is unlikely to grow.
However, if the radiative feedback causes significant disruption to the flow, then
falling densities within $\rdust{}$ are likely to allow the ionised volume to
grow, violating our assumptions.

\begin{table}
\centering
\begin{tabular}{ll}
Gas Density & \unit{10^{-16}}{\grampercubiccmnp} \\
Gas Velocity & \unit{\scinot{5}{5}}{\cmpersecnp} \\
Stellar Mass & \unit{10}{\Msol}
\end{tabular}
\caption{Chosen parameters for Bondi--Hoyle simulations}
\label{tbl:ChoseParamsBHsims}
\end{table}


\section{Dynamical Simulations}
\label{sec:dynamics}

The Bondi--Hoyle solution does not occur for gaseous
flow.
Shocks and instabilities occur, requiring numerical
solutions.
To simulate the flow, we used the latested publically available
version (2.0.3) of the \Zeus{} code of
\citet{1992ApJS...80..753S} in spherical polar mode.
Radiative feedback was included using the radiative
transfer algorithm developed by \citet{2003MNRAS.338..962E},
with the radiative feedback appearing as an extra source of
momentum.
In the present work, we solved the diffusion approximation (and
hence $\rdust{}$ as well) along each
ray using the assumption of spherical symmetry.
This should be a good approximation on the upstream side, but will
be poor in the wake, where the density gradient will tend to transport
radiation sideways.
This would have the effect of expanding the wake, and reducing its ram
pressure.
Despite the computational efficiency of our new algorithm, it is
still expensive to solve on every timestep.
Accordingly, the density structure along each ray was stored whenever
the radiative feedback was calculated.
The solution for a particular ray was only recomputed when this
structure changed signficantly.

We calculated accretion by depleting the innermost grid cells
of material, by a fixed fraction each timestep.
On the upstream side, the boundary conditions were those given
by equations~\ref{eq:BHanalyticvr} to~\ref{eq:BHanalyticrho}.
This enabled the outer boundary of the simulation to be moved inwards
(cf \citet{1991MNRAS.252..473K}), easing the computational load.
On the downstream side, we imposed outflow conditions.
For the equation of state, we tried $\gamma = 1, 4/3, 5/3$.
Although the densities suggest isothermality is the best approximation
\citep{1976MNRAS.176..367L}, there will also be radiative heating
and possibly other processes.
The star was parameterised by the ZAMS formul\ae{} of
\citet{1996MNRAS.281..257T}.

In the majority of our calculations, we used 50 grid cells in
the $\theta$ direction, and 160 in the radial direction.
The angular spacing was uniform, but the radial spacing was not.
Only the inner five radial zones (extending to $r=0$) were
uniformly spaced, and formed the accretion cavity.
The remaining 155 grid cells were logarithmically spaced, joining
smoothly on to the uniform grid.
For some test cases, we increased the resolution, but this did not
change our results significantly.
The resolution we used means that most of the direct stellar
impulse is applied in a single grid cell, but \Zeus{} does not
seem to mind this (cf section~\ref{sec:DirectOnly}).
In the isothermal case, the accretion column is covered by about
five azimuthal zones.

In all our calculations, the only source of gravity is the
star itself.
The self-gravity of the gas is neglected, due to the computational
cost.
However, the neglect is not without justification.
The gas mass contained in our simulations should remain less than
the stellar mass.
Furthermore, these simulations are meant to be a small sample
around a larger cluster core.
In this case, the structure of the cluster itself would provide
some support against local gravitational collapses of the gas.
As we shall see, isothermal gas can give rise to high
gas densities in the accretion column, and it is
possible that self-gravitating clumps could form.
However, such clumps would still be accreted by the star.
Finally, any over-densities which do form will tend to absorb
more radiation, and the increased temperature will tend to stabilise
them against gravitational collapse.

\subsection{Flow without Feedback}

As mentioned above, the ratio $\zetaHL{}/\rdust{}$ is rather large
for our simulations - larger than most simulations of Bondi--Hoyle
flow to date.
Therefore, we first simulated Bondi--Hoyle flow without feedback, to
provide a benchmark for the runs with feedback included.

\begin{figure}
\centering
\includegraphics[scale=0.5]{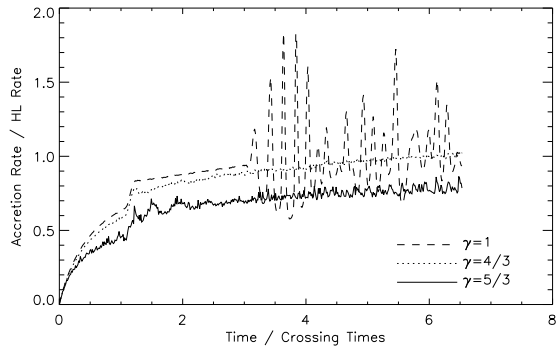}
\caption{Accretion rates for plain Bondi--Hoyle flow}
\label{fig:PlainBHflowMdotRun91011}
\end{figure}

Figure~\ref{fig:PlainBHflowMdotRun91011} shows the resultant accretion
rates as a function of time.
The simulations with $\gamma$ values of $4/3$ and $5/3$ settle into
a quasi-steady state, with the stiffer equation of state giving
a slightly lower accretion rate.
The isothermal run shows quite strong oscillations in its accretion
rate, although the mean value is close to $\mdotHL{}$.
These oscillations are likely to be due to the gas suddenly shocking
as it encounters the $\theta = 0$ axis.
In a very short distance, it has to turn around (the test of
accretion in the standard analysis is for material flowing
\emph{away} from the accretor), and it is evidently unable to
do this in a steady manner.

\begin{figure}
\centering
\begin{tabular}{c}
\includegraphics[scale=0.5]{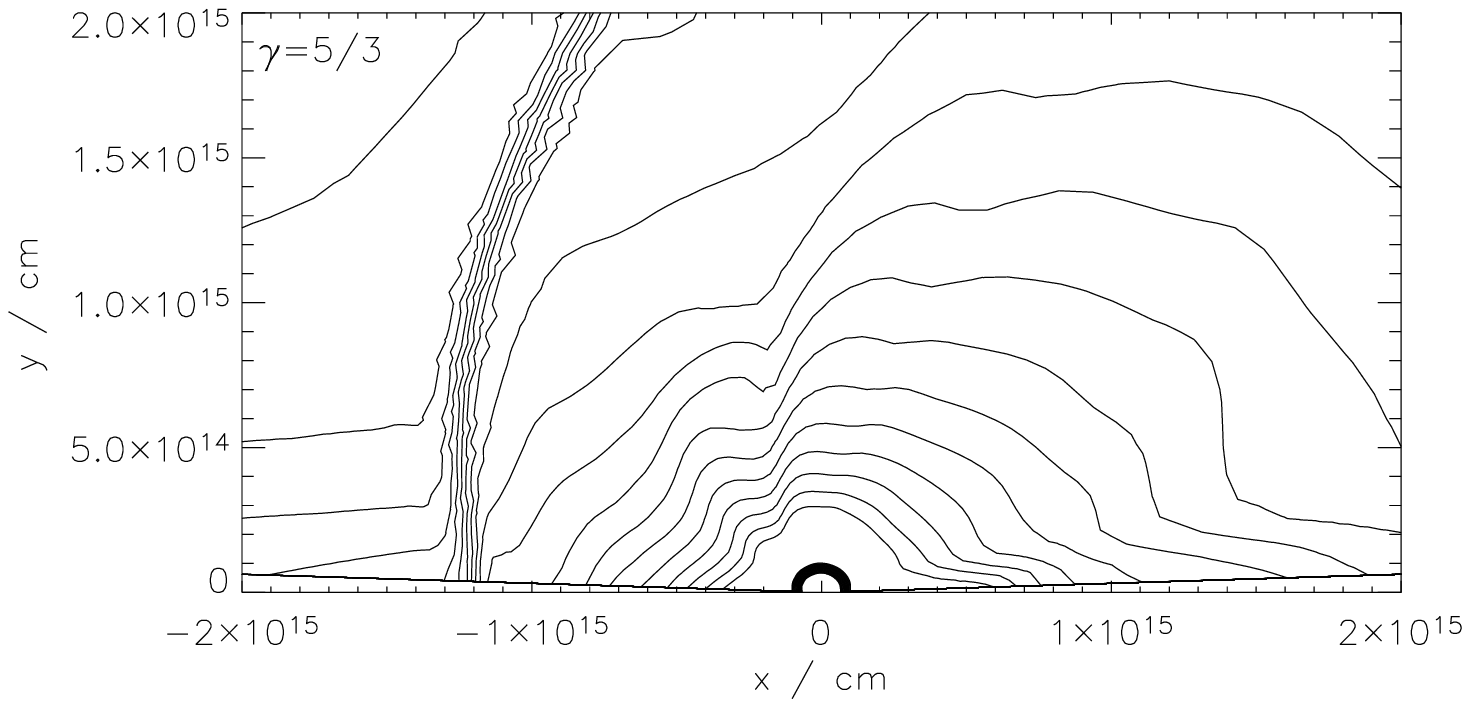} \\
\includegraphics[scale=0.5]{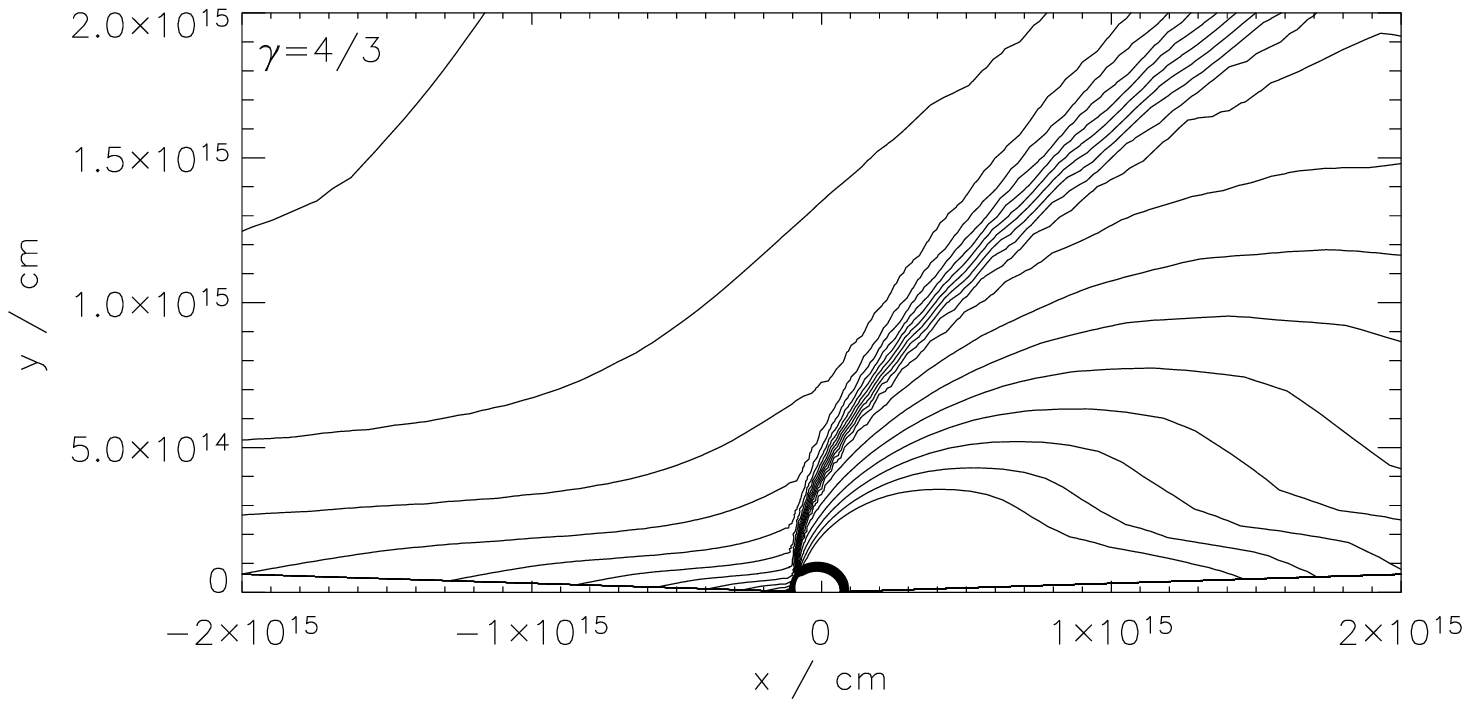} \\
\includegraphics[scale=0.5]{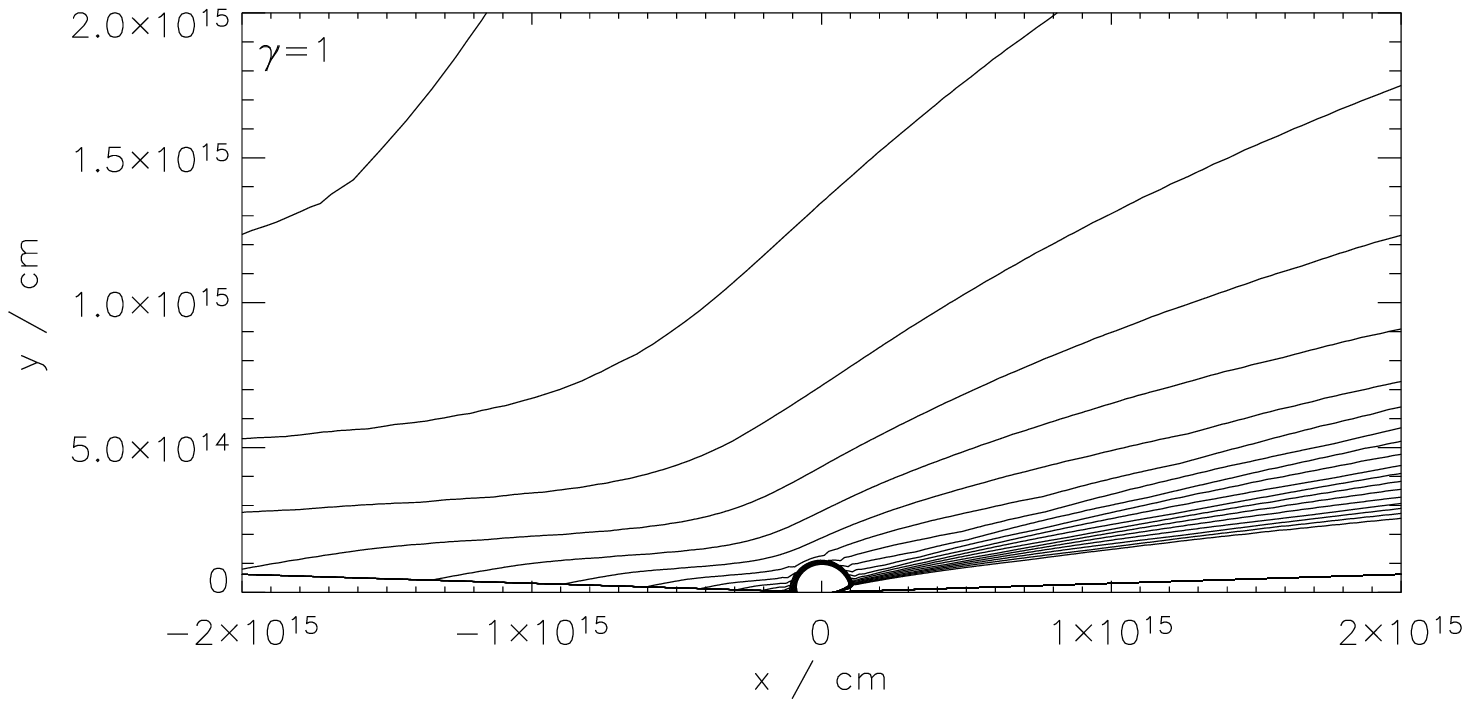}
\end{tabular}
\caption{Density contours for high Mach number flow around a small accretor}
\label{fig:PlainBHflowDensityContoursRun91011}
\end{figure}

\begin{figure}
\centering
\begin{tabular}{c}
\includegraphics[scale=0.5]{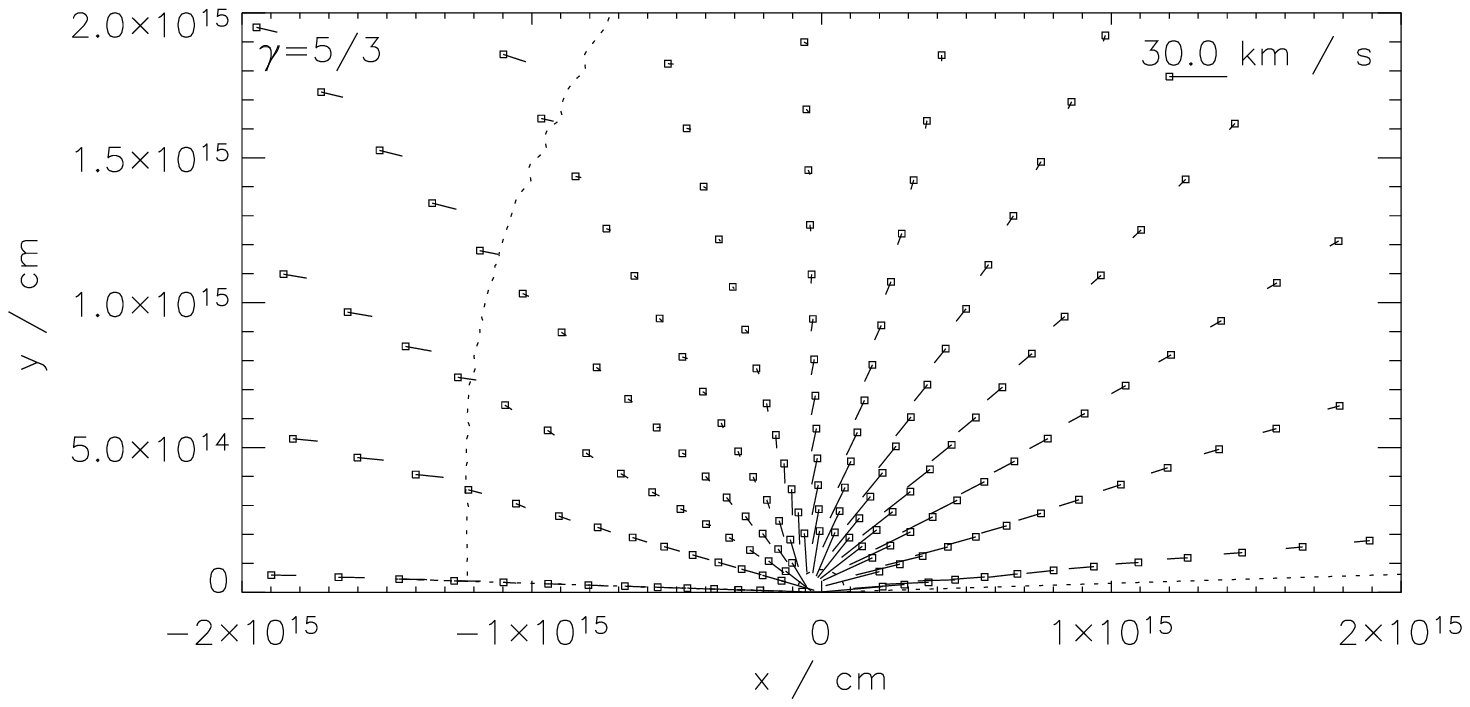} \\
\includegraphics[scale=0.5]{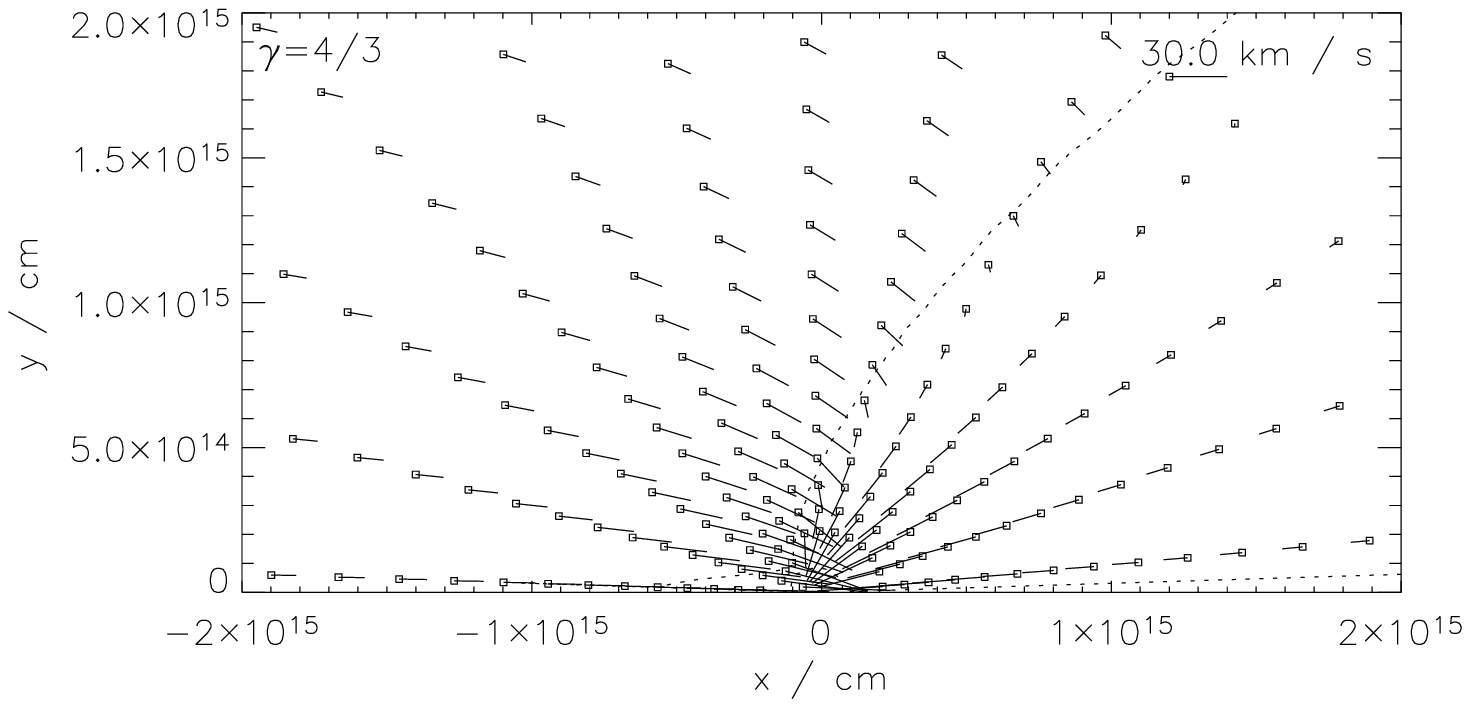} \\
\includegraphics[scale=0.5]{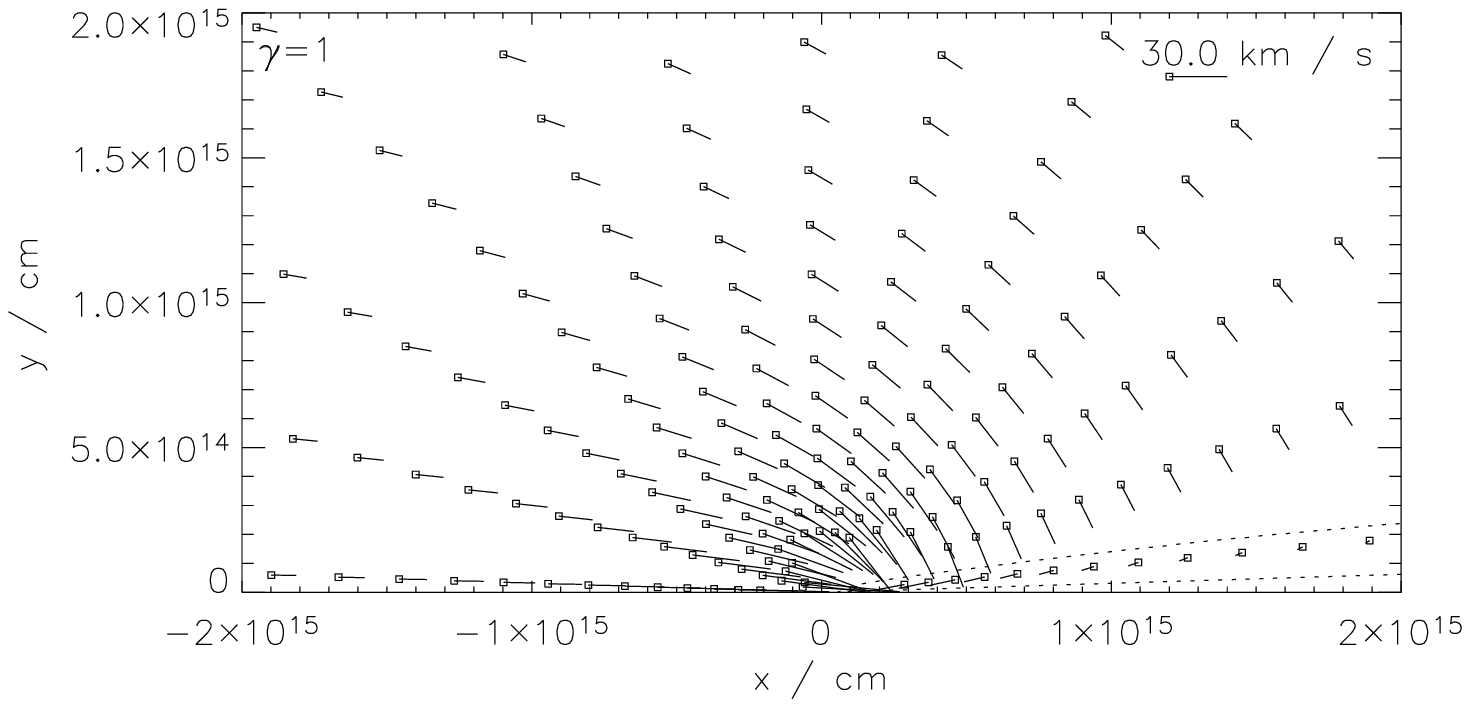}
\end{tabular}
\caption{Velocity fields for high Mach number flow around a small accretor.
The approximate location of the shock front is shown by the dotted line.
Flow is incident from the left}
\label{fig:PlainBHflowVelFieldsRun91011}
\end{figure}

We plot sample density and velocity fields for these runs
in Figures~\ref{fig:PlainBHflowDensityContoursRun91011}
and~\ref{fig:PlainBHflowVelFieldsRun91011}.
Note that the size of these plots is rather smaller than
$\zetaHL{} \approx \unit{10^{16}}{\centi\metre}$.
While the isothermal run is obviously qualitatively similar
to the flow described by \citeauthor{1939PCPS.34..405}, the
stiffer equations of state lead to rather different behaviour.
For $\gamma=5/3$, a bow shock forms, with the downstream
flow close to spherical (cf e.g. \citet{1971MNRAS.154..141H}).
Putting $\gamma=4/3$ moves the bow shock back, attaching it
to the accretor.
The isothermal flow is qualitatively similar to that
described by equations~\ref{eq:BHanalyticvr}
to~\ref{eq:BHanalyticrho}.
We found that the oscillating accretion rate plotted in
Figure~\ref{fig:PlainBHflowMdotRun91011} was due to a slight
expansion and contraction in the accretion column, rather
than a dramatic change in the flow pattern.

\begin{figure}
\centering
\includegraphics[scale=0.5]{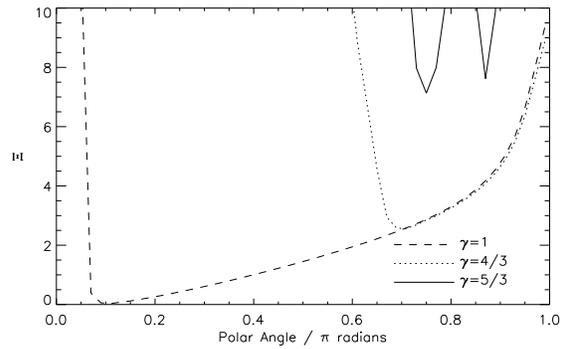}
\caption{Ratio of radial ram pressure to $L/c$ for high Mach number flow around a small accretor}
\label{fig:PlainBHflowAccelRatioRun91011}
\end{figure}

We can also make a useful comparison between the ram pressure in the
radial direction and the sharp impulse at the dust destruction
radius.
At the most basic level, accretion can only proceed if
\begin{equation}
\Xi = \frac{4 \pi \rho \vdust{2} \rdust{2}}{\frac{L}{c}} > 1
\label{eq:XiAccRatioDefine}
\end{equation}
(the angular factors are the same for both ram and radiation
pressure).
For a \unit{10}{\Msol} star, the formulae of
\citet{1996MNRAS.281..257T} predict
$L = \unit{5552}{\Lsol}$.
Results for an expected dust destruction radius of
\unit{\scinot{2.1}{14}}{\centi\metre} are plotted in
Figure~\ref{fig:PlainBHflowAccelRatioRun91011}.
From this, it is obvious that we have happened upon an
interesting region of parameter space.
The adiabatic runs have $\Xi$ values which get as
low as a few.
This means that the direct impulse will not be sufficient
to halt the flow.
However, the extra momentum from the thermalised field
may slow the flow sufficiently to allow the accretion
flow to be shut down.
The $\Xi$ values for the isothermal case vary even more
dramatically.
Just outside the wake, the radiation pressure at the
dust destruction front will dominate the ram pressure
(unsurprising, given the velocity field shown in
Figure~\ref{fig:PlainBHflowVelFieldsRun91011}).
In the wake, the ram pressure is orders of magnitude greater
than the radiation pressure.

\subsection{Direct Impulse Only}
\label{sec:DirectOnly}

\begin{figure}
\centering
\begin{tabular}{c}
\includegraphics[scale=0.5]{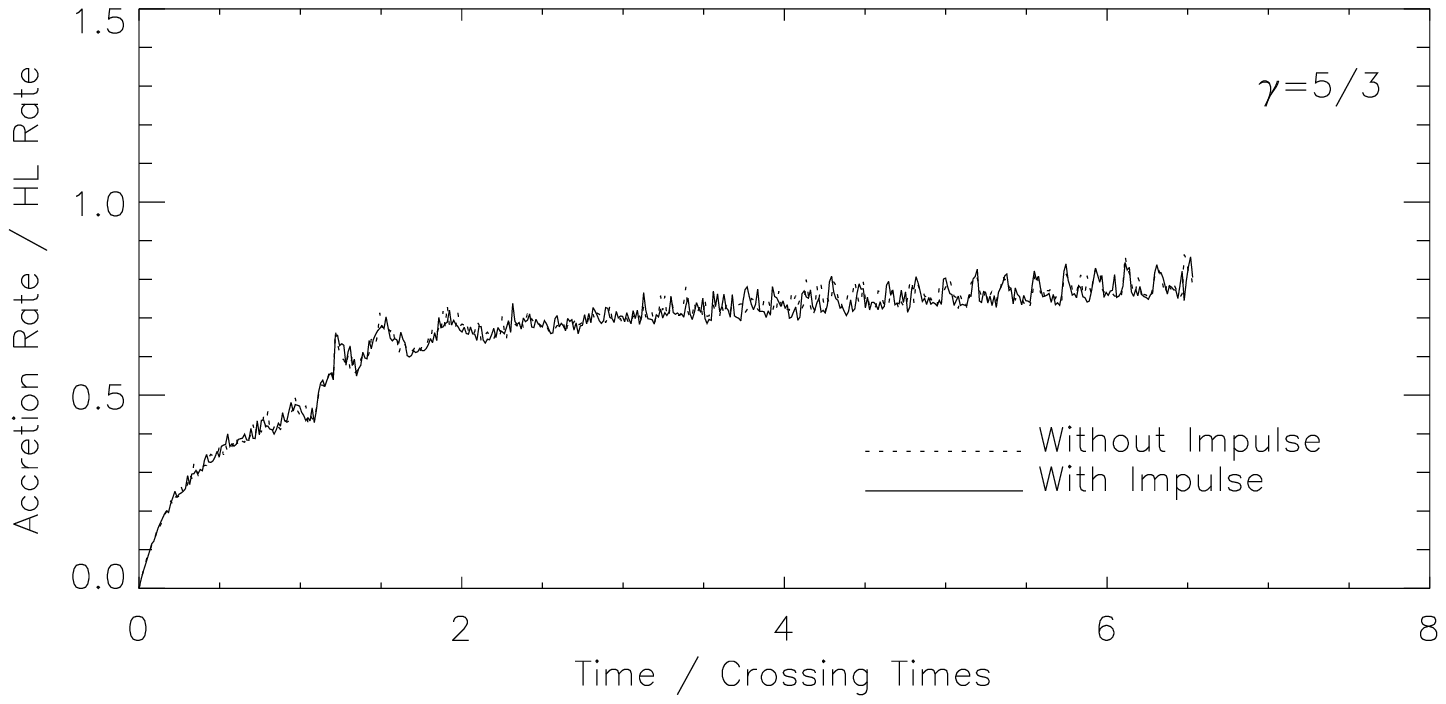} \\
\includegraphics[scale=0.5]{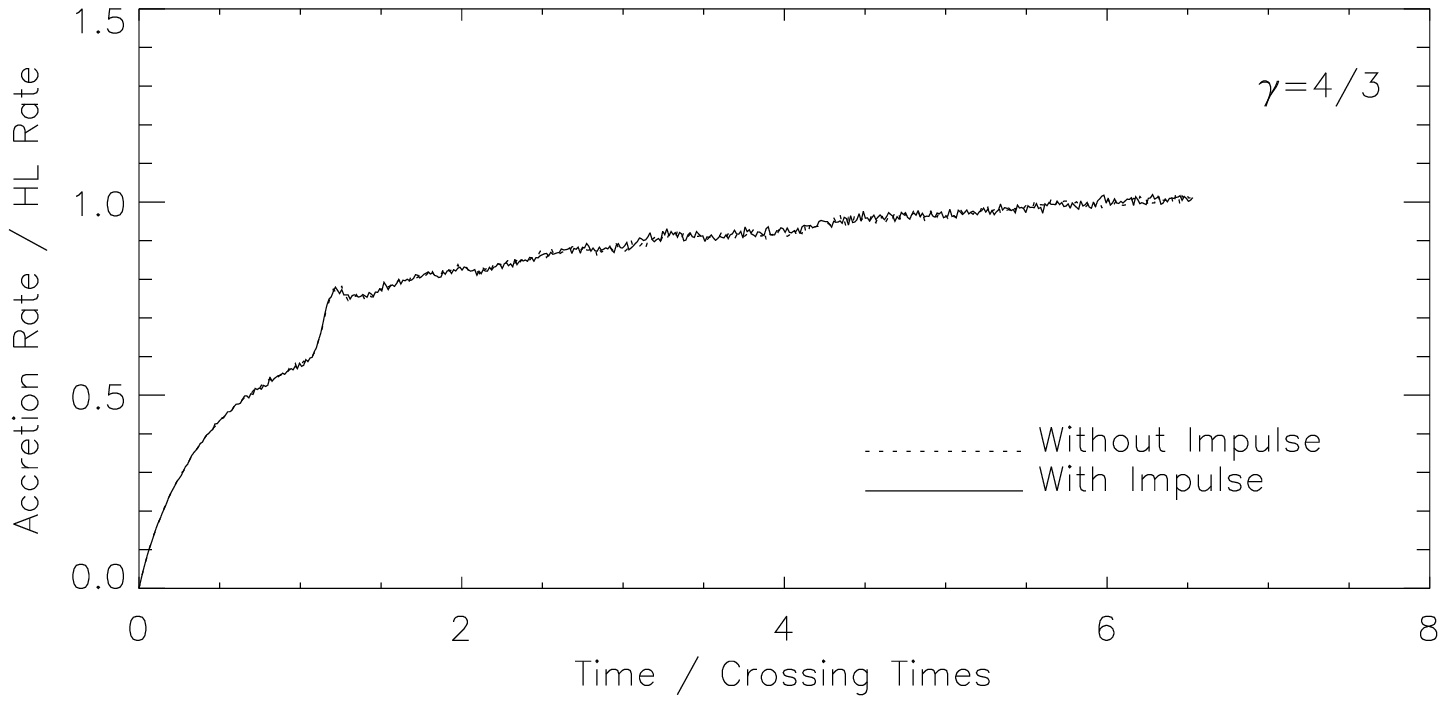} \\
\includegraphics[scale=0.5]{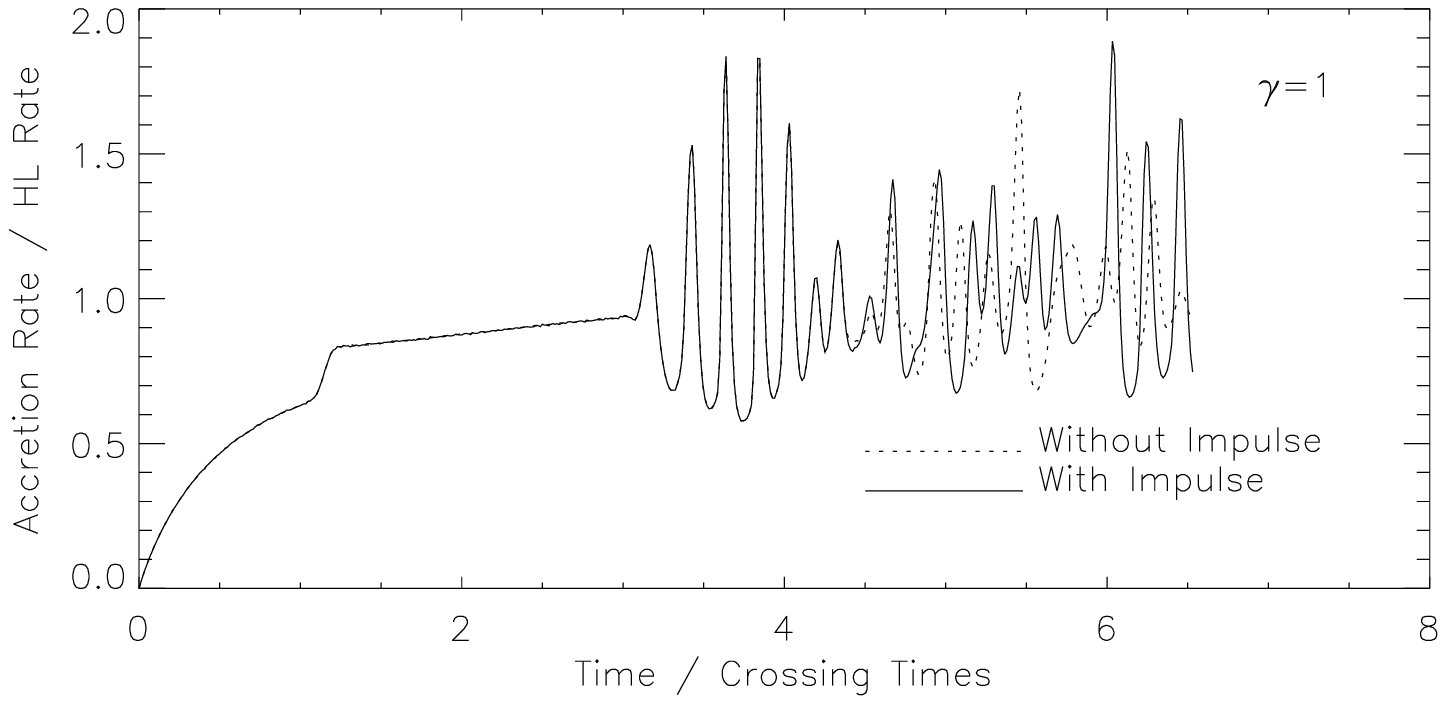}
\end{tabular}
\caption{Accretion rates for runs with the direct impulse only.
The corresponding accretion rates from
figure~\ref{fig:PlainBHflowMdotRun91011} are plotted for
comparison}
\label{fig:SimpleBHMdot}
\end{figure}

To provide comparison with later simulations, we performed three runs
with only the direct impulse included.
We achieved this by injecting $L/c$ worth of momentum into the grid at
a fixed radius.
As may be seen from figure~\ref{fig:SimpleBHMdot}, the effect
on the accretion rate is minimal.
The greatest differences occur for the isothermal case, and
even then, the changes probably have more to do with the intrinsic
instability of the flow, rather than any significant changes
due to feedback.
These results are not too surprising in the light of
figure~\ref{fig:PlainBHflowAccelRatioRun91011}, which showed that
the ram pressure should exceed the radiation pressure for most
of the flow at the dust destruction radius.

\subsection{Feedback with Reduced Dust}

Three runs were performed with the depleted dust abundances of
\citet{1987ApJ...319..850W}.
This involves an ad hoc reduction of the dust abundances by
a factor of ten, and also a reduction in graphite grain size.
\citeauthor{1987ApJ...319..850W} found this reduction was
necessary to ensure material was always accelerating inwards
in their calculations.

\begin{figure}
\centering
\begin{tabular}{c}
\includegraphics[scale=0.5]{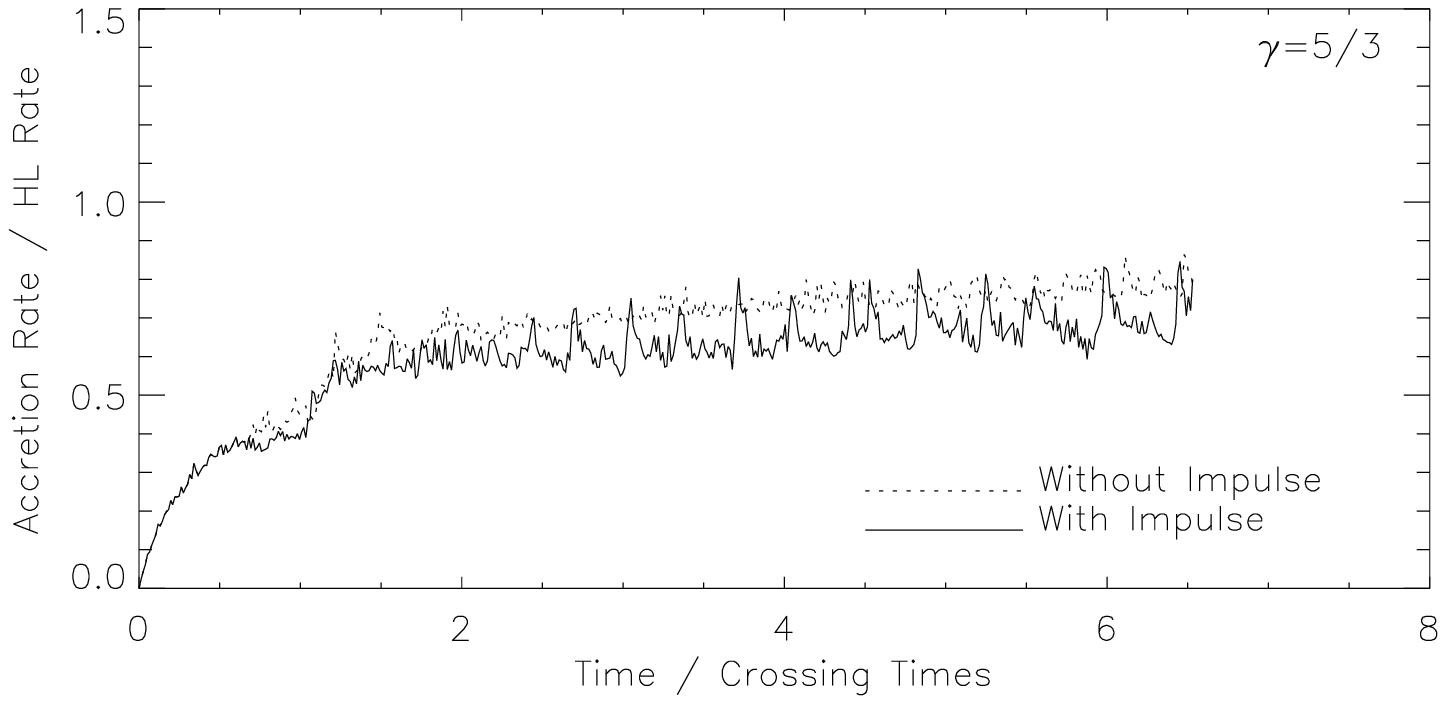} \\
\includegraphics[scale=0.5]{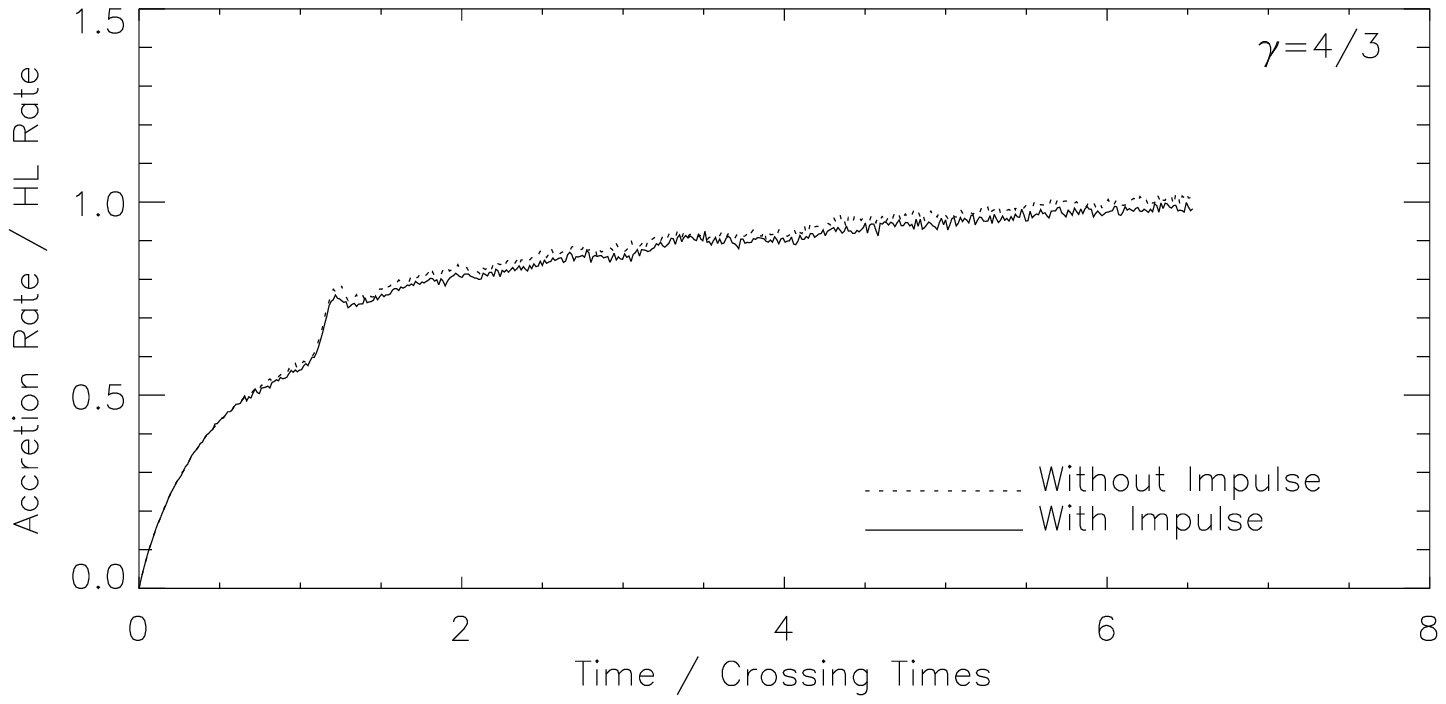} \\
\includegraphics[scale=0.5]{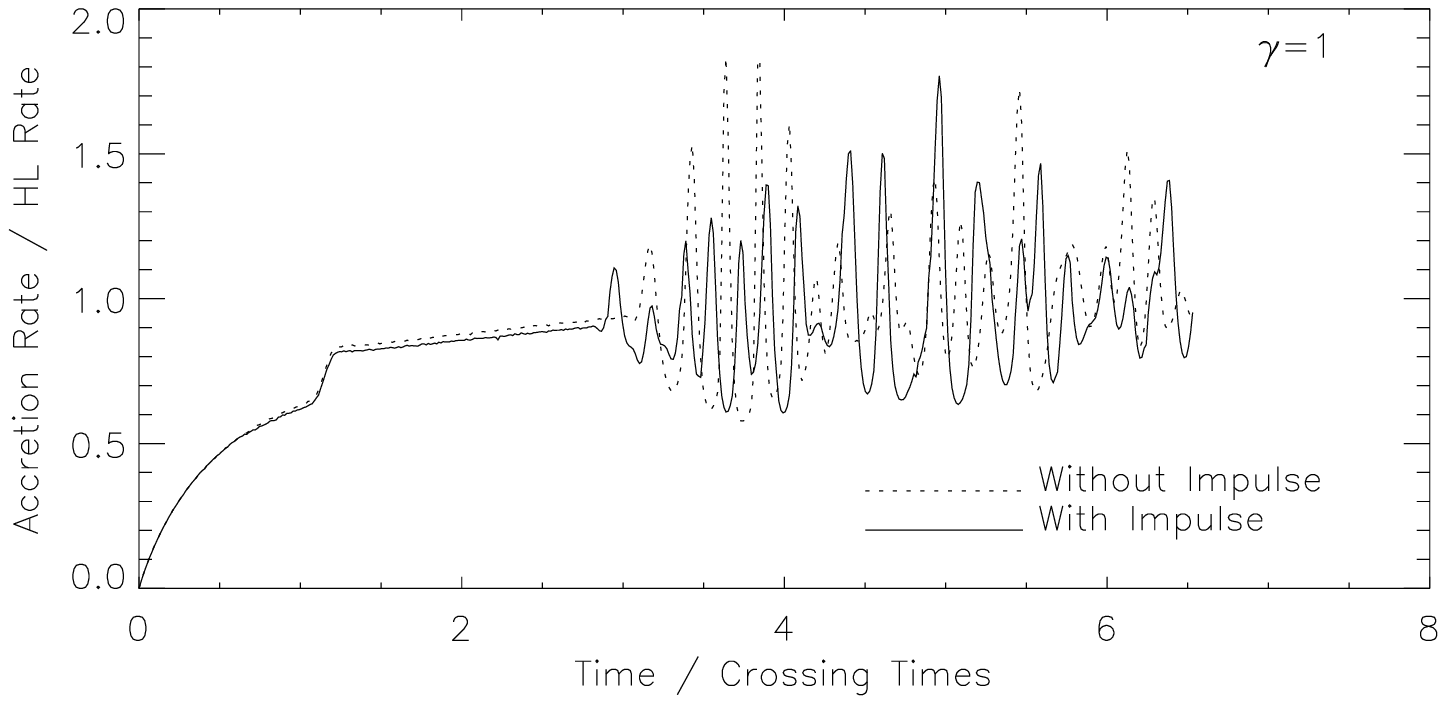}
\end{tabular}
\caption{Accretion rates for runs with reduced dust.
The corresponding accretion rates from
figure~\ref{fig:PlainBHflowMdotRun91011} are plotted for
comparison}
\label{fig:ClothoReduceDustMdot}
\end{figure}

Accretion rates for these runs are plotted against time in
figure~\ref{fig:ClothoReduceDustMdot}.
From these, it is apparent that the addition of feedback is
not having a dramatic effect on the flow.
The run with $\gamma=5/3$ now shows larger fluctuations in its
accretion rate, and the overall rate has dropped too.
Smaller changes have occurred for the $\gamma=4/3$ gas, with
the accretion rate only falling slightly when compared to the
pure hydrodynamic run.
For the isothermal case, the change in accretion rate is barely
visible compared to the large fluctuations present.

On examining the density contours, we found that the run with
$\gamma=4/3$ was little changed - unsurprisingly, in view of
Figure~\ref{fig:ClothoReduceDustMdot}.
The isothermal simulation showed a small density enhancement
adjacent to the wake, corresponding to the portion of
Figure~\ref{fig:PlainBHflowAccelRatioRun91011} where $\Xi$ was
less than unity.
Beyond this, there were few changes.
There were greater changes for the $\gamma=5/3$ gas - the flow
was far less smooth (as one might expect, given the changing accretion
rate).
Since the reduced dust abundances made the Rosseland optical depth
low, the shape of the dust destruction front was close to
spherical in all cases.
The only significant deviation was for the accretion column in
the isothermal case.
Even here, the dust destruction front was only pushed out
slightly.

\subsection{Feedback with Full Dust}

\begin{figure}
\centering
\includegraphics[scale=0.5]{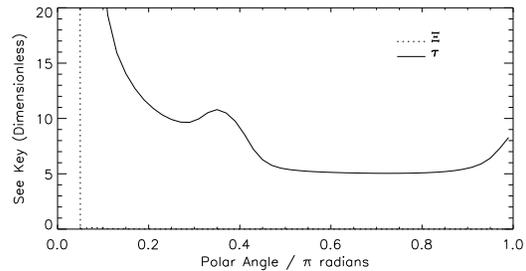}
\caption{Comparison between the Rosseland optical depth ($\tau$) and
the ratio of ram and radiation pressures ($\Xi$) for the isothermal
run with full dust}
\label{fig:ClothoFullDustXiTauCompareIso}
\end{figure}

Increasing the dust abundances to their normal galactic values had
a dramatic effect on the flow.
Soon after the impulse switched on, material ceased penetrating
$\rdust{}$, and the grid cells inside drained of material.
\Zeus{} was then forced to terminate.
Consider Figure~\ref{fig:ClothoFullDustXiTauCompareIso}.
This plots $\Xi$ and $\tauR{}$ values for the isothermal
gas shortly after the impulse is switched on.
If the gas is outflowing, $\Xi$ is set to zero.
It is relevant to compare these two quantities because the amount
of momentum imparted to the flow by the thermalised field is
$L \tauR{} / c$.
Since $\Xi$ measures the ram pressure of the flow at $\rdust{}$ in
units of $L/c$, if $\tauR{} \gg \Xi$, then the thermalised radiation
should reduce the ram pressure at $\rdust{}$ enough to allow the
direct radiation field to halt inflow.
Figure~\ref{fig:ClothoFullDustXiTauCompareIso} shows that the total momentum
imparted to the flow by the luminosity completely dominating the ram pressure
in the flow.
This is the reason accretion halted.

When we examined the position of the dust destruction front, we found
that it was still close to spherical for most of the flow.
However, the attached shock in the $\gamma=4/3$ case, and the
accretion column in the isothermal case caused noticeable
deviations.
In the isothermal run, the dust destruction front in the
accretion column was located roughly four times further
from the accretor than on the upstream side.

We attempted various numerical `fixes' in order to ensure that
numerical problems were not interfering with the physical
analysis.
In some runs, we spread the effect of the direct stellar radiation
field over a number (typically seven) of grid cells.
In others, we tried adjusting the CFL condition, to incorporate
a term based on $s = 0.5 a t^2$.
Finally, we tried both methods together.
None of these `worked' - accretion was still halted in every
case.
We therefore concluded that our results were not an artefact of
the numerical method.
This is consistent with the theoretical arguments discussed
above.


\section{Discussion}
\label{sec:discussion}

Radiative feedback has a much more dramatic effect on
Bondi--Hoyle accretion than spherically symmetric
accretion.
This is brought about by the lower central concentration
and presence of centrifugal support found in Bondi--Hoyle
flow.

Although we have only been able to study one set of parameters,
there are several implications for the formation of massive
stars in clusters.
Since $L \sim M^{3}$ for stars close to \unit{10}{\Msol}, the fact that our
simulations were obviously on the borderline between accreting and not
accreting implies that
\begin{itemize}
\item Less massive stars will have few problems accreting in the Bondi--Hoyle geometry
\item More massive stars are unlikely to accrete significant amount of mass under these conditions
\end{itemize}
Unfortunately, it is not easy to determine an exact limit - especially
since we have not been able to test a wide range of parameters, and the
fact that it is the reprocessed radiation which is of critical importance.
Consider equation~\ref{eq:BHshockVelChange}.
Since $L / \rdust{2}$ should be approximately constant
(cf the Stefan--Boltzmann equation), if the
`inside' radial velocity is positive (indicating accretion) for a protostar at some mass,
it should remain positive as the mass increases.
The sharp impulse at the dust destruction radius cannot stop the inflow by itself.
It can only do so if the flow has been slowed by the thermalised radiation.
Of course, if the value of $\vinf{}$ rises too much (due to the contraction of the cluster),
then the accretion timescale (cf equation~\ref{eq:HoyleLyttletonAccRateDefine})
will become longer than the main sequence timescale of the star.

Estimating the effect of the thermalised radiation is much more difficult, since
the impulse from it is distributed throughout the flow.
Based on the findings presented above, requiring that
$\Xi > \tauR{}$\footnote{Recall that $\Xi$ is the ratio of ram to radiation pressure
at $\rdust{}$ - see equation~\ref{eq:XiAccRatioDefine}} at the dust destruction radius is
the most reasonable condition for the purpose of making estimates.
However, this should be treated with caution, since it is not rigorously
derived.

\begin{figure}
\center
\includegraphics[scale=0.5]{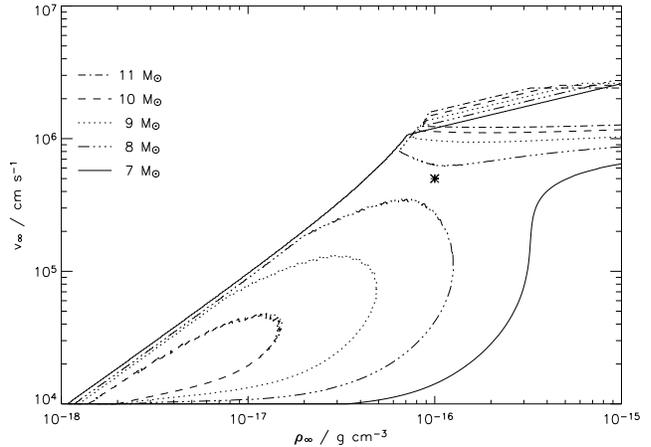}
\caption{Allowed parameters for Bondi--Hoyle accretion for a variety of
stellar masses, as suggested by the conditions in the text.
The contours enclose the allowed regions for the marked stellar masses, when the
dust abundance is assumed to be Galactic.
The star marks the location of the simulations presented in this paper}
\label{fig:AllowedRegions}
\end{figure}

Despite these problems, we can still suggest regions of parameter space which
might permit accretion, and hence be worthy of future investigation.
These regions must fulfil several conditions:
\begin{itemize}
\item $\mdotHL{}$ must be large enough to permit the star to gain appreciable mass
      before it leaves the main sequence

\item They must have $\Xi > 1$ at the dust destruction radius (or the flow will be
      reversed)

\item The \HII{} region must lie within the dust destruction radius

\item They must have $\Xi > \tauR{}$ at the dust destruction radius (again, to avoid
      reversal of the flow)
\end{itemize}
For a given mass of star, these conditions define a region of $(\rhoinf{},\vinf{})$
space which are likely to permit accretion in the Bondi--Hoyle geometry.
In Figure~\ref{fig:AllowedRegions}, we show the allowed parameters for a variety of
stellar masses,
evaluated using equations~\ref{eq:BHanalyticvr} to~\ref{eq:BHanalyticrho} for
the full dust abundances.
For the first condition, we required $\mdotHL{} > \unit{10^{-5}}{\Msol\usk\yyear}$.
The last three conditions above are functions of angle, so in these cases
we tested the condition on the upstream ($\theta=\pi$) direction.
The \HII{} region will be largest in this direction.
Furthermore, the analytic approximation will hold to smaller radii.
We also added the condition that $\zetaHL{} > 2 \rdust{}$, and took the edge of the
calculation to lie at $2 \zetaHL{}$ (for the purposes of calculating $\tauR{}$).
Both of these conditions are somewhat arbitrary.
First, note that the parameters tested above
($\rhoinf{} = \unit{10^{-16}}{\grampercubiccmnp}$ and
$\vinf{}=\unit{\scinot{5}{5}}{\cmpersecnp}$ for $M = \unit{10}{\Msol}$)
lie outside the allowed range for the full dust (this point lies inside the
permitted region, as calculated for the reduced dust).
Of course, this is not a full test of the validity of these conditions, but it is
encouraging.
The sloping straight line boundary is caused by the requirement on $\dot{M}$.
The horizontal boundary which appears close to
$\vinf{} = \unit{\scinot{2}{7}}{\cmpersecnp}$ for the higher masses is set by
the constraint on $\zetaHL{}$.

For accretion to occur, Figure~\ref{fig:AllowedRegions} suggests that
the optical depth must be kept low, either by high velocities (so that the distance
to the outer boundary is small) or low densities.
The permitted region for the higher mass stars at high velocities is almost
independent of density because
$\rho$ does not vary much between $\rdust{}$ and $2 \zetaHL{}$.
This means that $\rho$ cancels from the ratio $\Xi / \tauR{}$, leaving the ratio a
function of $\vinf{}$ only.
At low densities, the rather complicated shape of the allowed region is due to the
solution of the thermalised radiation field, and the calculation of $\tauR{}$.
Note that there is no allowed low density region for the \unit{11}{\Msol}
case.

It seems that Bondi--Hoyle accretion
is going to run into serious trouble at about \unit{10}{\Msol}.
Even for masses of only \unit{8}{\Msol}, the allowed region is curtailed
quite severely.
However, these are only estimates, and assume that Bondi--Hoyle accretion
will be a good approximation over all the parameter space covered by
figure~\ref{fig:AllowedRegions}.
Unfortunately, this is unlikely to be the case.
For example, a velocity of \unit{10^4}{\cmpersecnp} for a \unit{10}{\Msol} star
gives $\zetaHL{} \sim \unit{10^{19}}{\centi\metre}$, which is far larger than the
interstellar separation in a cluster core.\footnote{Additionally, simulating such a large
volume would be computationally painful, due to the enormous difference in length scales.
We will still have $\rdust{} \sim \unit{10^{14}}{\centi\metre}$.
Worse still, a cloud that big is going to be self-gravitating}
Also, a \unit{10}{\Msol} star moving at \unit{10^5}{\cmpersecnp} through at cloud
with $\rhoinf{} = \unit{10^{-15}}{\grampercubiccmnp}$ would have an unreasonably
high accretion rate of $\mdotHL{} \sim \unit{0.3}{\Msol\usk\reciprocal\yyear}$.
These concerns would restrict the allowed parameter space further.

There are further problems, which our simulations do not address.
We have not included an accretion luminosity, since it will complicate the feedback
mechanism.
However, the accretion luminosity implied by the parameters of Table~\ref{tbl:ChoseParamsBHsims}
is roughly three times the ZAMS luminosity.
A short test, on the assumption that the accretion luminosity was also isotropic
(this may not be the case), suggested that accretion was not possible even for the
reduced dust abundances.
Our inability to simulate the ionised region is also a concern.
Flow in the Bondi--Hoyle geometry is far less centrally concentrated than in
the spherically symmetric case, so the \HII{} region might be able to
grow.
If it can reach the dust destruction radius, some rather unpleasant chemistry
would have to be simulated as well.
Our imposition of axisymmetry and low resolution in the inner portions of the grid
also prevent us examining the question of accretion discs.
If the flow at infinity is non-uniform (rather likely in a protocluster), then
there is the \emph{potential} for angular momentum accretion too
\citep{1997A&A...317..793R,1999A&A...346..861R}.
This would drive the formation of an accretion disc.

The problems radiative feedback presents in the Bondi--Hoyle geometry have
potentially serious implications for the merger model of massive
star formation proposed by \citet{1998MNRAS.298...93B}.
Accretion drove the contraction of their clusters to the high stellar densities
required for mergers.
\citet{2002MNRAS.336..659B} presented a full hydrodynamic simulation of such
a cluster, but did not have a cut off mass for accretion (their model was
scale free).
If the most massive stars in the cluster core stopped accreting, then it is
possible that the cluster core would not contract enough to drive a high merger rate.
However, this is very uncertain.
\citet{2001MNRAS.323..785B} simulated another cluster, and their results
show many low mass stars in the core, in addition to higher mass ones.
It is possible that the lower mass stars could accrete enough material
to drive the stellar density to that required for mergers.
Note also that \citet{1998MNRAS.298...93B} did prevent stars more massive
than \unit{10}{\Msol} accreting in their simulations, but still found merging
stars.


\section{Conclusion}
\label{sec:conclude}

We have shown that if stars form in the ultra-dense cores of young
clusters, as postulated in the models of
\citet{1998MNRAS.298...93B,2002MNRAS.336..659B}, then the feedback
from radiation pressure on dust is likely to prevent accretion on
to stars more massive than $\sim \unit{10}{\Msol}$.
This feedback is more disruptive to the accretion flow than
in our previous simulations \citep{2003MNRAS.338..962E} which used
the initial conditions employed in the classic work of
\citet{1987ApJ...319..850W} (i.e. spherically symmetric accretion and
initial gas densities of \unit{10^{-19}}{\grampercubiccmnp}).    

The reasons for this more effective feedback in the present case
are two-fold.
Firstly, the ultra-dense cluster cores that are invoked by
\citeauthor{1998MNRAS.298...93B}are unobservably short lived.
In this phase the local gas density of their cores is extremely high,
exceeding that seen in regions of massive star formation, or that
employed by \citeauthor{1987ApJ...319..850W}, by many orders of
magnitude.
Consequently, the optical depth to the thermalised radiation field
($\tauR{}$) is very high.
Since the total rate of momentum input into the gas is
$\sim \tauR{} L/c$, this increases the coupling between the radiation
and gas dramatically.
Secondly, in the case of spherically symmetric  accretion, the gas
density is centrally concentrated and the velocity field is radial.
Both of these effects enhancing the radial component of the ram pressure
in the accretion flow at the dust destruction radius.
In the case of  cluster cores, by contrast, accretion proceeds in
Bondi--Hoyle geometry, and flow is nearly tangential as it approaches
the accretion wake (see Figure~\ref{fig:BHgeometry}).
Consequently the radial component of the ram pressure is relatively
small, and this again favours the disruption of the accretion flow. 

The above results imply that \emph{if} massive stars indeed form in
these hypothetical ultradense cores, then for masses greater than
$\sim \unit{10}{\Msol}$, they must be assembled entirely by stellar
collisions and consequent mergers rather than by accretion.
The attainment of these ultradense conditions is however driven by
accretion by stars in the core and it currently  unclear whether
accretion on to low mass stars (i.e. those with mass $< \unit{10}{\Msol}$)
will suffice for this purpose. 
This issue  needs to  be explored by further simulations that incorporate
feedback from radiation pressure on dust into realistic cluster
simulations.


\section*{Acknowledgments}

The authors would like to thank Jim Pringle and Ian Bonnell for several useful discussions.
RGE is also particularly grateful to Matthew Bate and Jim Stone, for their aid with {\sc ZEUS}.
An anonymous referee also made several helpful suggestions.


\bibliography{bibs/radiativetrans,bibs/dust,bibs/zeus,bibs/observations,bibs/hydro,bibs/compute,bibs/stellarevolve,bibs/accretiondisks,bibs/starform,bibs/bondihoyle}
\bibliographystyle{mn2e}

\bsp

\label{lastpage}

\end{document}